\magnification=\magstep1
\hsize=15truecm
\vsize=22truecm
\hoffset=0.5truecm
\def\double{\baselineskip=0.7truecm}

\raggedbottom
\font\text=cmr10
\text
\def\deg{\hbox{$^\circ$}}
\def\ltsim{\mathrel{\hbox{\rlap{\hbox{\lower4pt\hbox{$\sim$}}}\hbox{$<$}}}}
\def\gtsim{\mathrel{\hbox{\rlap{\hbox{\lower4pt\hbox{$\sim$}}}\hbox{$>$}}}}


\def\apj    {{\it Ap.~J.~}}
\def\apjl   {{\it Ap.~J.~(Letters)~}}
\def\apjs   {{\it Ap.~J.~Suppl.~}}

\def\aj     {{\it A.~J.~}}
\def\aa     {{\it Astr.~Ap.~}}
\def\mnras  {{\it M.N.R.A.S.~}}
\def\aas    {{\it Astr.~Ap.~Suppl.~}}
\def\pasp   {{\it P.A.S.P.~}}
\def\araa   {{\it A.R.A.A.~}}

\def\nature {{\it Nature~}}

\def\ref {\noindent} 
\def\v#1  {{\bf {#1}~}}

\def\etal {{\it et al.~}}

\double

\centerline{\bf HST Observations of the Field Star Population}
\centerline {\bf in the Large Magellanic Cloud}

\vskip 1cm

\centerline { Rebecca A. W. Elson$^1$, Gerard F. Gilmore$^2$, \& 
Basilio X. Santiago$^3$}

\vskip 1cm

\centerline{\it $^1$ Institute of Astronomy, Madingley Rd.,Cambridge CB3 0HA, UK} 
\centerline{\it email: elson@ast.cam.ac.uk} 

\centerline{\it $^2$ Institute of Astronomy, Madingley Rd.,Cambridge CB3 0HA, UK}
\centerline{\it  email: gil@ast.cam.ac.uk} 

\centerline{\it $^3$ Federal Rio Grande do Sul, Universidade, Instituto de F\'isica} 
\centerline{\it 91501-970 Porto Alegre, RS Brasil} 
\centerline {\it email: santiago@if.ufrgs.br}

\vskip 4cm
\noindent
Submitted to \mnras  13 January 1997

\noindent
Revised: 4 March 1997

\vskip 3cm
\noindent
$^1$ Based on data obtained as part of the {\it Medium Deep Survey}

\vfill\eject

\centerline{\bf Abstract}

We present  $V$ and $I$ photometry, obtained with the Hubble Space
Telescope, for $\sim
15,800$ stars in a field in the inner disk of the Large Magellanic
Cloud.  We confirm previous results indicating that an intense star
formation event, probably corresponding to the formation of the LMC
disk, occurred 
a few times $10^9$ years ago.  We find a small but
real difference between our field and one further out in the disk observed
by Gallagher \etal (1996): either star formation in
the inner disk commenced slightly earlier, or the stars are slightly
more metal rich.  We also find evidence for a later burst, around
1 Gyr ago, which may correspond to the formation of the LMC
bar.  About 5\% of the stars in our field are 
substantially older than either burst, and are probably members of an old
disk or halo population with age $\sim 12$ Gyr.

\noindent
{\bf Keywords: Galaxies : Stellar Content : Evolution : Magellanic Clouds}

\vskip 1cm
\noindent
{\bf 1. Introduction}

Deep surveys of faint galaxies have revealed an ``excess'' of faint 
blue objects and it has been suggested that these
may be dwarf galaxies at redshifts $z\ltsim 0.5$
(lookback times of $\ltsim 7$ Gyr) 
undergoing intense bursts of star formation (cf. Cowie, Songalia \& Hu, 1991;
Babul \&
Ferguson 1996).  We have in our vicinity a number of dwarf galaxies known
to have undergone bursts of star formation (see Hodge 1989
for a review).  
Detailed studies of their star formation
history and field star populations can contribute to our understanding
of the evolution of such galaxies, of the mechanisms provoking 
bursts, of the time scales and mass functions
involved, and of what such galaxies look like prior to the onset of
the bursts.  The time scales and mass functions  
are particularly important as input for population synthesis
models which attempt to interpret the integrated light of galaxies 
too distant to resolve into stars.

Among our neighbours, the closest and best studied is the Large Magellanic
Cloud (LMC), believed to have undergone a global burst of star formation
a few Gyr ago, possibly provoked by a close encounter with the Milky Way.
The LMC is still experiencing a significant amount of star formation, which
gives it a patchy, irregular appearance, but its underlying light is 
well represented by an exponential disk with scale length 
$\alpha = 0.65$ deg$^{-1}$ ($=0.74$ kpc$^{-1}$)
(cf. de Vaucouleurs 1960; Elson, Fall \& Freeman 1987).  
The inner $\sim5\deg$ of the disk, imaged in 
the Hodge-Wright Atlas (Hodge \& Wright 1967), is relatively well studied.
The disk extends to at least $\sim9\deg$, 
and its rotation curve appears to  be constant out
to this distance, and possibly to $\sim 15\deg$, 
suggesting the presence of a dark matter halo
(cf. Schommer \etal 1992).
 
Until recently, the rich star clusters 
in the LMC have provided the primary
tracers of its star formation history (cf. Girardi \etal 1995).  
The clusters range from a few million years old to
objects as old as the globular clusters in the halo of our own galaxy.
There appears to be a gap in the 
age distribution: a thorough investigation of the richest
clusters has so far yielded no examples with age 
between about 4 and 10 Gyr (cf. Da Costa, King \& Mould 1987; Jensen, Mould \& 
Reid 1988). Clusters with age less than $\sim 4$ Gyr share the kinematics
of the young disk.
The oldest clusters also appear to  
have disk-like, rather than halo-like  kinematics, although the sample
is small ($N\sim 10$) so the properties of this disk are poorly determined 
(Freeman, Illingworth \& Oemler
1983; Schommer \etal 1992).  

Field stars provide a more direct and detailed probe of the star formation
history of the LMC.
Recent ground-based
studies of the field star population have been carried out by Bertelli
\etal (1992), Westerlund, Linde, \& Lynga (1995),
and Vallenari \etal (1996a,b).  Bertelli \etal examine
three regions $\sim5\deg$ from the centre of the
LMC, with photometry to $V_0 \ltsim 23$.  Their data suggest a global
burst of star formation $2-4$ Gyr ago, with the exact timing depending 
on the choice of stellar evolution models.  Vallenari \etal use the
same method and similar data to investigate the stellar population in
six fields, and reach a similar conclusion.  They also suggest that
the time of onset of the burst varies with location. 
Westerlund \etal study regions in the NW and SW of the LMC, and find a
strong component with age $1-3$ Gyr, and a very weak component with age
$7-10$ Gyr in each region.
Gallagher \etal (1996) use data from the Hubble Space Telescope (HST) 
in the
$V$ and $I$ bands to construct a colour-magnitude diagram (CMD) 
to $V \sim 25$ for 2000 stars in a
field $\sim 4\deg$ from the centre of the LMC. 
They conclude that a peak in the star formation
rate occurred $\sim 2$ Gyr ago, and that there was little star formation
activity prior to $\sim 8$ Gyr ago.  

While these studies have focussed on the intermediate age population,
others have addressed the question of what the LMC looked like prior to the 
onset of the burst, and whether it has an extended stellar halo.  The
oldest globular clusters, as already mentioned, have kinematics which are
not consistent with a pressure supported halo. 
Old long-period variables have kinematics consistent with a thick
disk or flattened spheroid, with velocity dispersion $\sim 33$ km/s and 
scale height $\sim 0.8$ kpc (Hughes, Wood \& Reid 1991). 
Other tracers of a Population II component in the LMC are
giants which are too blue to 
be of intermediate age, 
and RRLyraes (Kinman \etal 1991).  
Giants too blue to belong to an intermediate age population are
present in many CMDs of field stars used to 
subtract the background
in studies of star clusters.  Their numbers 
are generally small, and they have 
rarely been commented on, although occasionally they have been cited
as evidence of an older population (cf. Da Costa \etal 1987; Westerlund
\etal 1995).
Field RRLyraes have been studied by Kinman
\etal who
conclude from a sample of 80 objects that the LMC has a Population II
halo or thick disk extending to $\sim17\deg$ and 
containing about 2\% of the mass of the LMC, a fraction similar to
that in the stellar halo of the Galaxy.
For a review of the intermediate and older populations in the LMC,
see Olszewski, Suntzeff, \& Mateo (1996).
 
Finally, the third important component of the 
stellar population in the LMC is the bar, which is $\sim 2\deg$ long,
and is offset from the centre of the disk.
Ground based observations of the bar are severely limited by crowding, so
little is known of when it formed or how long it took to do so.
Hardy \etal (1984) studied a field near the NW end of the  bar with
photometry to 
$V\sim 21$,  and found that the bulk of star formation there began
earlier than  $\sim 1$ Gyr ago. 
Bica, Claria \& Dottori (1992) 
have used the star clusters in the bar to
attempt to trace its evolution.  They report an absence of clusters
with ages $2-6$ Gyr, suggesting that the bar formed $\ltsim 2$
Gyr ago, however their results 
are uncertain because of the relatively small number of clusters, and because
of the possibility that some of the clusters are merely seen superposed
against the bar, and are not actually associated with it.

To summarize, a picture is emerging of this ``irregular'' galaxy  as composed of
a Population II halo or thick disk, an intermediate age  
disk that formed quite quickly
$2-4$ Gyr ago, and a bar which formed a few Gyr after that.  In this
paper we use data from the HST
to attempt to place further limits on the properties of these
three components.
In Section 2 we present HST observations in $V$ and $I$ of 
a field $\sim 1.3\deg$  from the centre of the LMC,   
acquired as part of the Medium Deep Survey.  Our photometry 
extends to $V_0 \sim 25$, 
well below the main-sequence turn-off of a $\sim 10^{10}$ year old population.
The contribution to our sample of an old and intermediate age
population arepopulation are  analysed in Section 3. 
Our results are summarized and discussed 
in Section 4.

\vskip 1cm
\noindent
{\bf 2. Observations}

Our field is located  at RA=05:35:36, $\delta$=$-$69:25:36 (J2000),
near the SE end of the LMC bar, $0.7 \deg$ off the
major axis of the bar, and $1.3 \deg$ from the kinematic centre of the
LMC.  It lies just 2 arcmin west of the association NGC 2050.
It was observed with WFPC2 on board HST on 
1996 January 30.  The F814W ($\sim I$) and F606W ($\sim V$) filters were used.
Two exposures of 500 seconds were taken in each of the filters.
The field size is 4.56 arcmin$^2$ (excluding the unexposed borders
of each chip).  The fwhm is $\sim 1.5$ pixels, 
and the image 
scale is 0.1 arcsec pixel$^{-1}$.  To ensure uniform star 
selection and photometry,
we did not use the smaller Planetary Camera image, which would not increase our
sample size significantly.

The images were reduced using the standard pipeline procedure,
and coadded.  In order to remove cosmic ray events, the smaller of
the two pixel values at each position in each chip was adopted.
We used the DAOPHOT task DAOFIND to identify all sources in the images
to a threshold of 5 $\sigma$ in the F814W image.  
We constructed a PSF from isolated stars
in each chip, and
fit this to all the sources detected. In addition
to stars, DAOFIND
identifies a significant number of spurious detections, particularly along
diffraction spikes and surrounding saturated stars.  To eliminate these
from our sample, we used the
parameters ``sharpness'' and $\chi^2$, which characterize the goodness of
fit of the PSF to each object.  We plotted each against magnitude, and 
eliminated outlying objects. (General details of this approach are presented in
the context of a study of the globular cluster $\omega$Cen by Elson
\etal 1995). 

Aperture photometry in the undersampled WFC
images can give more accurate magnitudes than PSF fitting,
as judged by the width of the resulting main-sequence; this is 
particularly true in fields where isolated, well-exposed stars for 
constructing a PSF are difficult to find.
We constructed CMDs from the PSF magnitudes and from magnitudes derived
from aperture photometry with an aperture with radius 2 pixels.  Indeed, the 
main-sequence was $\sim 25\%$ narrower in the CMD constructed
from aperture photometry. We therefore adopted magnitudes from 
aperture photometry 
rather than from PSF fitting.

We applied the following 
corrections, to our 2-pixel aperture magnitudes,
following Holtzman \etal (1995a,b):  (1) A small
correction for geometric distortion;  this was always less than
0.04 mag, and independent of filter. 
(2) Aperture corrections
to a 10 pixel radius which contains 100\% of the light.  These were 
0.26 mag in $V_{606}$ and 0.28 mag in $I_{814}$, 
from Table 2b of Holtzman (1995a); 
uncertainties are $\pm 0.05$ mag.  (3) Synthetic zero points
from Table 9 of Holtzman \etal (1995b).  
(4) A reddening correction of $E(B-V)=0.20\pm0.04$ 
from Hill, Madore \& Freedman (1994), which was 
determined from $UBV$ photometry of 72 stars in 
the association NGC 2050, 
just 2 arcmin east of our field.  This corresponds to
$E(V-I)=0.27$ and $A_V=0.63$ (Taylor 1986).  
We note that any variations in the redenning in this area of the
LMC on a scale $\ltsim 2$ arcmin will contribute to the uncertainties
in our results.  However, the agreement with the results of 
Gallagher \etal (1996), discussed in Section 3.1 below, gives us confidnece
in our adopted value of redenning.
Finally, we adopted a distance modulus for the LMC of $18.5 \pm 0.1$,
derived by Panagia \etal (1991) from SN1987A. 
This corresponds to a distance to the LMC of 50.1$\pm 3.1$ kpc.
At this distance, $1\deg = 0.875$ kpc.

A dereddened instrumental CMD for the $\sim 15,800$ stars in our sample 
is shown in Figure 1.  We converted the instrumental magnitudes to the 
Johnson-Cousins system using synthetic calibrations taken from Table 10
of Holtzman \etal (1995b).  CMDs in the Johnson-Cousins system are shown
in Figures 2a-c. Stars
brighter than $V_0\sim 18.6$ are saturated. 
Representative Poisson error bars are shown, and the additional
systematic uncertainties from sources mentioned above are $\delta V_0=
\delta I_0 = \pm 0.16$, and $\delta (V-I)_0 = 0.23$.
Tables with photometry are available electronically from R. Elson.

\vskip 1cm
\noindent
{\bf 3. Stellar Populations}

To facillitate analysis of our CMD, we constructed $(V-I)_0$
histograms in $0.5$ mag
bins to define the loci of the main sequence and giant
branch.  The histograms are shown in Figs. 3a-l.
The colours of the peaks were estimated by eye.  Fitting Gaussians gives similar
results in some cases, but in many cases (eg. the blue edge of the 
upper main-sequence with its tail redwards) a Gaussian function is
not appropriate.
The peak value(s) of $(V-I)_0$ for each bin are plotted as filled
circles in Figs. 2a-c.  We have checked that the positions of these
peaks are not sensitive to the choice of binning. 

Superposed on the CMDs are Yale isochrones
(Green, Demarque \& King 1987) with $Y=0.3$ and a range of metallicities
and ages.  We based our choice of metallicities on
the results of Olszewski \etal (1991)
who plot [Fe/H] vs age for 29 LMC clusters with metallicities determined  
from sepctra of a total of 80 individual stars.
The value $Z=0.01$ is appropriate for populations with age $\ltsim$
1 Gyr, while the values $Z=0.01$ and $Z=0.004$  
bracket the likely metallicities of populations with age 
$1-5$ Gyr.  For a population with age $\gtsim 5$ Gyr the range 
$0.004 < Z < 0.0001$ is appropriate. 

The main features of our CMD are as follows:

The upper main sequence is well represented by
the $Z=0.01$ ZAMS.  Its redward extension 
indicates a range of stellar ages, but
because our image is saturated at $V_0 \approx 18.6$, we are unable
to say anything quantitative about recent star formation in this field.

There
are two peaks in the colour distribution of intermediate age stars
with $20 < V_0 < 23$; these peaks
are clearly visible in the histograms in Figs. 3c-g, and are suggestive
of two distinct main-sequence turnoffs in the CMD, and hence of
two distinct bursts of star formation.

The lower main-sequence at $V_0 > 23$ is best represented by the
isochrones with $Z=0.004$, although
at $V_0\sim 23$ the loci of the data are
slightly bluer than these isochrones.
This may indicate the presence of a more metal poor population as suggested 
by the isochrones with $Z < 0.001$ in Fig. 2c.  

The bulk of stars on the giant branch are significantly
bluer than the intermediate age isochrones, regardless of metallicity,
but are well represented by the oldest isochrones.  There is a small but 
significant number of redder giants,  well represented by the intermediate
isochrones.  

There is  a
prominent red horizontal branch ``clump" of giants just below our 
saturation limit,
at $V_0 \sim 19$.  Studies show that the 
magninitude and colour of this clump are
essentially independent
of the age of the population for stars with ages $\gtsim 0.5$ Gyr (Hardy
\etal 1984; Hatzidimitriou 1991).  

The overall features of this comparison between our CMD and the isochrones 
are not dependent on the choice of isochrones;  for example,
the Bertelli 
\etal (1994) isochrones give the same general results.
We now discuss in detail the different populations represented in
our sample.

\medskip
\noindent
{\bf 3.1 The Intermediate Age Populations:  Formation of the Disk and Bar?}

The $(V-I)_0$  histograms in Figs. 3c-g show two clear peaks 
(not including the reddest one corresponding to the giant
branch, and in (c), the bluest one, which reflects the most
recent star formation). 
These must correspond to two populations with distinct ages.  
Comparison of the loci of the stellar distribution
with the isochrones in Figs. 2a and b suggests an age
for the younger population of 
slightly less than 1 Gyr, or of $\sim 2$ Gyr, 
depending on the adopted metallicity.  The redder peak corresponds to 
a population with age $\sim 2-4$ Gyr,
again depending on metallicity.  

Given the 
proximity of our field to the bar, we tentatively identify the younger 
component with the bar population.  Our age estimate agrees well
with those of Hardy \etal and Bica \etal discussed above.
The age of the older component corresponds well with age estimates
for the widely
studied burst of star formation 
at $2-4$ Gyr.  We identify this global burst with the
formation of the disk of the LMC.  A more precise dating of the onset and
duration of the burst is difficult because such estimates are sensitive
to the adopted (range of) metallicity, and to the choice of stellar 
isochrones, and because the number of stars belonging unambiguously to
any one age range is small.

While absolute measurements are difficult, relative measurements are 
more straightforward.  If, for instance, we are correct in 
identifying the two intermediate populations with the disk and bar,
then whatever the absolute ages, the bar formed $\sim 1-2$ Gyr after
the disk.  It would be interesting to determine from models whether
an instability in a disk with the observed properties of the LMC 
disk would produce such a bar on this time scale.  In this context it
is important to determine the age of the LMC bar 
more  precisely, as it is the only such structure close enough to be
resolved into stars, and therefore for which the star formation history
and mass function can be studied in detail.

Relative measurements can also reveal 
whether there are spatial variations in the
age and/or metallicity of the intermediate age disk population as claimed,
for example, by   
Vallenari \etal (1996b).
Strong variations, particularly with position angle, are unlikely:
at a radius of $\sim 4 \deg$ the
orbital velocity in the LMC is $\sim 60$ km s$^{-1}$ (Hughes \etal 1991;
Schommer \etal 1992),
and a typical intermediate age star will have orbited the LMC about 4 times.  
With a velocity dispersion in the disk of $\sim 5-10$ km s$^{-1}$
(Hughes \etal 1991), a star
2 Gyr old will have travelled $10-20$ kpc in a random direction.
The intermediate population 
should therefore be well mixed, and a sample drawn from
any one position will contain stars that formed at random positions
in the disk.  
  
To explore whether our data contain any evidence for 
age or metallicity gradients in the 
disk, we compare our CMD with that of Gallagher \etal (1996)
which was obtained with the same instrument, filters, and calibrations
as ours, for
a field $\sim 4 \deg$ from the centre of the LMC, near the cluster 
NGC 1866.  
Figures 4a-f
show a comparison between the $(V-I)_0$ histograms that Gallagher \etal derive
for their sample, and those for our sample, for five different magnitude bins.
In the two brightest bins there are small differences reflecting the 
variations in recent star formation between the two locations.
The absence in the Gallagher \etal sample 
of the red peak from the giant branch in Fig. 4a is due simply to
the smaller size of their sample ($\sim 2000$ stars compared to our 
$\sim 15,800$ stars).  In the intermediate bins
at  $V_0\approx 21.1$ and $V_0\approx 21.7$  
in Figs. 4c and d, our histograms are significantly
redder than those of Gallagher \etal  
The peaks are plotted in the CMDs in Figs. 5a-c,
with isochrones with two different metallicities superposed. 
Either our (inner) field is slightly more 
metal rich (Fig. 5a), or star formation in our field commenced slightly earlier
than in the outer disk (Figs. 5b and c), or both.
Olszewski \etal find evidence of a slight metallicity difference, in
this same sense, between
clusters closer and further than 5$\deg$ from the centre of the
LMC.  

The histograms for the faintest magnitude bin, shown 
in Fig. 4e are in excellent agreement, the only difference
being that, with the longer exposure times of Gallagher \etal
their errors are smaller and their colour distribution
is therefore narrower.
There is no offset in colour between the two samples.  This further
justifies our adopted value of reddening.  If, for example, we had adopted
a smaller reddening of $E(B-V)=0.1$, the giant branches of the isochrones in 
Figs. 2a
and b would pass closer to the bulk of the giants, and the main sequence
would still fit adequately (though would be too
blue at the faintest end).
However, there would be an offset in intrinsic colour
between our sample and that of
Gallagher \etal in the faintest bin,
as illustrated in Fig. 4f, while the difference in the intermediate magnitude
bins would essentially vanish.  It
would be difficult to explain such a large difference in intrinsic
colour of main-sequence stars
at faint magnitudes without a corresponding difference at slightly brighter
magnitudes (Figs. 4c and d).

Finally, we compare the main-sequence 
stellar luminosity function in our field with those in
other parts of the LMC disk. 
Figure 6 shows main-sequence luminosity functions for our field, and
for those from Bertelli \etal and Vallenari \etal corrected for incompleteness.
In our field, stars blueward of the dashed line in Fig. 2a were selected as
main-sequence stars.
No completeness corrections are required for our data at $M_V < 4$. 
We have normalized the luminosity functions to
coincide at $M_V=3$ ($V_0=21.5$), below the turnoff of the intermediate
age population.  Brighter than $M_V\sim3$ there are differences among
the luminosity functions, which reflect variations in the recent star 
formation history across the LMC disk.  Fainter than $M_V\sim 3$ the slopes
of the luminosity functions appear to be quite similar, suggesting that
there are no large
differences in initial mass function among the 
different fields, although deeper observations are required to confirm
this (cf. Santiago \etal 1997). 

\medskip
\noindent
{\bf 3.2 A Population II Halo or Thick Disk?}

A picture is now emerging of the LMC in which an exponential disk formed
$\sim 2-4$ Gyr ago, followed by a bar which formed perhaps $\sim 1-2$ Gyr 
after that.  But what did the LMC look like before the formation of the
disk and bar?  The rotation curve suggests the presence of a halo of 
dark matter (Schommer \etal 1992), and studies of 
about a dozen old globular clusters, $\sim 80$ RRLyrae
stars, and 63 old long period variables provide
some evidence for a Population II component.  
A few studies of normal field stars have also hinted at the presence
of such a population (cf. Da Costa \etal 1987; Elson, Forbes \& Gilmore 1994).
Neither the clusters nor the 
variables have the kinematics of an isothermal halo, but instead appear
to form a thick disk, although in the case of the clusters, the number
is so small that their systemic properties are uncertain.  
Kinman \etal explore the 
distribution of the RRLyraes and conclude that they are distributed 
either as a King model with tidal radius 
$r_t\sim 15 $ kpc, or in a disk with exponential
scale length $\alpha \sim 0.39$ kpc$^{-1}$ (compared to 0.74 kpc$^{-1}$
for the young/intermediate
disk).

An old stellar component is clearly visible in our 
CMD (Fig. 2c), in the giants just below the clump.  These giants are
too blue to be of intermediate age, and are
well represented by $12-15$ Gyr isochrones with $Z=0.0001$.  (As
illustrated, even a
7 Gyr isochrone has a giant branch too red to represent 
this population.)  For
comparison, the giant branches of two old globular clusters
in the LMC, NGC 1841 and NGC 2210, are also shown (Brocato \etal 1996).
These clusters have ages determined from deep CMDs to be
as old as the globular clusters in the halo
of our Galaxy.

To quantify the spatial distribution of this old stellar population,
we focus on the giant branch stars in a bin 
just below
the bottom edge of the clump, illustrated by the box 
in Fig. 2b.  These stars have $19.1 < V_0 < 20.1$ and $0.75 < (V-I)_0
< 0.95$.  The numbers of stars
in this box in our CMD and in that of Gallagher \etal is 60 and 12
respectively (in a field with area 4.56 arcmin$^2$).
These surface densities are plotted against distance ($R$)
from the centre of the LMC in Figs. 7a and b.  The adopted 
centre of the LMC is at RA=5:35:48.41, $\delta$=$-$69:25:16.12 (J1975).
The results of Kinman \etal 
for RRLyrae stars, normalized to overlap with the HST data, are also plotted. 
Figure 7a shows the King models which best fit the two HST points,
and the one derived by Kinman \etal for the RRLyraes.  
The King model preferred by Kinman \etal is clearly
inconsistent with our data, which indicates a higher central
surface density, if the giants and RRLyraes are indeed members of the 
same population.  
Figure 7b shows the same data plotted such that a 
straight line represents an exponential disk.  Again, our data and 
that of Kinman \etal suggest different scale lengths.  
Our results imply a central surface density of $15-20$
stars per arcmin$^2$
in the colour and magnitude range defined by the box in Fig. 2b.

The two HST  
points are clearly insufficient for further  
quantification of the structure of a Population II component of the
LMC, and we 
note only that the old, metal-poor component of the LMC 
includes a significant population of 
ordinary field giants as well as RRLyraes and globular clusters. 
In our CMD, the total number of Population II giants is $\sim 250$.
Assuming there is about the
same number of stars in this population at the turnoff, 
and assuming a Salpeter mass function below the turnoff,
we estimate that a total of $\sim 800$ stars in our CMD, or $\sim5\%$ 
of our sample,
are members of a population that substantially
predates the formation of the LMC disk.

Further quantification of the structure of this Population II halo/disk
must await the results of more deep imaging with HST, and analysis
of the field star variables discovered by the micro-lensing
surveys.  For example, Alard (1996) uses a stack of $\sim 60$ Schmidt 
plates to isolate a sample of $\sim 10,000$ variable stars in the LMC,
most of them RRLyraes.  The sample shows a strong central concentration
and appears to have a line-of-sight depth of $2-3$ kpc. 
A similar study of the RRLyraes in the outer parts of the LMC would
be valuable.

\vskip 1cm
\noindent
{\bf 4. Summary }

We have presented a CMD for $\sim 15,800$ stars in a field $\sim 1.3\deg$
from the centre 
of the LMC.  The following features are apparent in our sample:

(1) Ongoing star formation. 

(2) A population with age $\ltsim 1$ Gyr or $\sim 2$ Gyr, depending on
the adopted metallicity, which we tentatively identify with the LMC bar.

(3) A burst
of star formation $\sim 2-4$ Gyr ago, which we identify with the formation of 
the disk.  
Star formation either began slightly earlier in the inner disk, 
or the gas in the inner disk was 
slightly enriched compared to another
field at a radius of $\sim 4 \deg$.

(4) A population of stars ($\sim 5$\% of our sample) that substantially 
predates the formation of the disk. 

The star formation history in local 
dwarf galaxies is very varied, as illustrated
by Hodge (1989).  For example, recent studies of the Carina dwarf 
galaxy by Smecker-Hane \etal
(1996) show that it has a history of
episodic bursts and periods of quiescence, qualitatively similar to that
of the LMC. On the other hand, 
galaxies such as the Fornax dwarf elliptical appear to 
have consumed (or discarded) all their gas early in their evolution,
and are almost exclusively made of old stars.  
The explanation of the variation in star formation histories remains
elusive.  A common feature appears to be that all Galactic satellites
have survived very early star formation, with most retaining, or 
acquiring later, sufficient gas to sustain later star formation.  

Are these properties of the
local dwarf galaxies relevant to
understanding the
``excess'' faint blue galaxies observed
in deep surveys?  Cowie \etal (1991) determined redshifts for a 
small but complete sample of galaxies with $B=23-24$, and found
the sample to be dominated by objects with $z\sim 0.25$.  At this
redshift, assuming $H_0=75~ $km s$^{-1}$ Mpc$^{-1}$, these galaxies
would have $-17 \ltsim M_V \ltsim -19$, comparable to the present absolute
magnitudes of the Large and Small Magellanic Clouds ($M_V=-18.4$ and $-17.0$).
The LMC 
underwent a burst of star formation 
$\sim 2-4$ Gyr ago; Carina at 
$\sim 3-6$ Gyr ago, and $\sim 2$ Gyr ago.  
As lookback times these correspond to 
redshifts of $z\sim 0.1-0.4$.  Thus, a distant observer of the Milky Way
would observe its satellites flaring at these redshifts,
after which they would fade quite rapidly. 
The disappearance
of such satellites would be due to quiescence in their star formation,
and not to 
merging.  

The initial burst $\gtsim 12 $ Gyr ago in which the 
Population II component of such dwarf galaxies 
was formed would correspond to a redshift of 
$z\gtsim 2$.
Since the number of stars involved in the initial 
burst is small, these objects would be of low surface brightness, and prior to
$z\sim 0.5$ they would not be visible.
For example, the Fornax dwarf galaxy has $M_V\sim -13$.  At $z=0.5$ it would
have $V\sim 28$, which is near the limit of objects detected in the
Hubble Deep Field (cf. Mobasher \etal 1996).

An additional constraint on the evolution of the Galactic satellites was
derived by Unavane, Wyse \& Gilmore (1996), who compared the stellar 
populations of the Galactic field halo and the dwarf speroidal satellites
to show that at most a few large dwarf spheroidal satellites can have
merged into the galactic halo in the last $\sim 10$ Gyr.  We may
combine this constraint with the results of this paper, which strengthens
the evidence that dwarf satellites can survive several periods of
active star formation.  The combined evidence supports models of faint
blue galaxies in which galaxy number is approximately conserved, but
both total luminosity, and particularly surface brightness, evolve
considerably.

\vfill\eject
\noindent
{\bf References}

\ref
Alard, C. 1996 PhD Thesis, Universit\'e de Paris VI

\ref
Babul, A. \& Ferguson, H. 1996 \apj 458, 100

\ref
Bertelli, G. Bressan, A., Chiosi, C., Fagotto, F. \& Nasi, E. 1994 \aas 

106, 275

\ref
Bertelli, G., Mateo, M., Chiosi, C., \& Bressan A., 1992 \apj 388, 400

\ref
Bica, E., Claria, J. \& Dottori, H. 1992 \aj 103, 1859

\ref
Brocato, E., Castellani, V., Ferraro, F., Piersimoni, A. \& Testa, V., 1996

\mnras, in press

\ref
Cowie, L., Songalia, A. \& Hu, E., 1991 \nature 354, 460

\ref
Da Costa, G., King, C. \& Mould, J. 1987 \apj 321, 735

\ref
Elson, R., Fall, S. M. \& Freeman, K. 1987 \apj 323, 54

\ref 
Elson, R., Forbes, D. \& Gilmore, G. 1994 \pasp 106, 632

\ref
Elson, R., Gilmore, G., Santiago, B., \& Casertano, S. 1995 \aj 110, 682
 
\ref
Freeman, K., Illingworth, G. \& Oemler, A. 1983 \apj 272, 488

\ref
Gallagher, J. S. \etal 1996, \apj 466, 732

\ref
Girardi, L., Chiosi, C., Bertelli, G., Bressan, A., 1995 \aa 298, 87

\ref
Green, E.M, Demarque, P., \& King, C. R., 1987 Yale University Transactions,

Yale Univ. Observatory

\ref
Hardy, E., Buonanno, R., Corsi, C., Janes, K., \& Schommer, R. 1984 

\apj 278, 592 

\ref
Hatzidimitriou, D. 1991 \mnras 251, 545

\ref
Hill, R., Madore, B. \& Freedman, W. 1994 \apj 429, 192

\ref
Hodge, P. W. 1989 \araa 27, 139

\ref
Hodge, P. W. \& Wright, F. W. 1967 {\it The Large Magellanic Cloud}, Smithsonian

Press, Wahsington.

\ref
Holtzman, J. A., \etal 1995a  \pasp 107, 156

\ref
Holtzman, J. A., \etal 1995b  \pasp 107, 1065

\ref 
Hughes, S., Wood, P., \& Reid, N., 1991 \aj 101, 1304

\ref
Jensen, J., Mould, J. \& Reid, N. 1988 \apjs 67, 77

\ref
Kinman, T., Stryker, L., Hesser, J., Graham, J., Walker, A., Hazen, M.,

\& Nemec, J., 1991 \pasp 103, 1279

\ref
Mobasher, B., Rowan-Robinson, M., Gerogakakis, A. \& Eaton, N. 1996 

\mnras 282 7P

\ref
Olszewski, E., Schommer, R., Suntzeff, N. \& Harris, H. 1991 \aj 101, 515

\ref
Olszewski, E., Suntzeff, N. \& Mateo, M. 1996, \araa 34, 511

\ref
Panagia, N., Gilmozzi, R., Macchetto, F., Adorf, H.-M., \& Kirshner, R., 1991

\apjl 380, L23

\ref
Santiago, B., Elson, R., Sigurdsson, S., \& Gilmore, G. 1997, \mnras submitted

\ref 
Schommer, R., Suntzeff, N., Olszewski, E., \& Harris, H. 1992 \aj 103, 447

\ref
Smecker-Hane, T., Stetson, P., Hesser, J. \& VandenBerg, D. 1996 in ASP

Conf. Ser. 98, {\it From Stars to Galaxies}, eds. C. Leitherer, U. Fritze-v.

Alvensleben, \& J. Huchra, p. 328

\ref
Taylor, B. J. 1986 \apjs 60, 577

\ref
Unavane, M., Wyse, R. \& Gilmore, G. 1996 \mnras 278, 727

\ref
Vallenari, A. , Chiosi, C., Bertelli, G., Ortolani, S. 1996a \aa 309, 358 

\ref
Vallenari, A. , Chiosi, C., Bertelli, G., Aparacio, A. \& Ortolani, S. 1996b \aa 

309, 367

\ref
de Vaucouleurs, G. 1960 \apj 131, 574

\ref
Westerlund, B. E., Linde, P., \& Lynga, G. 1995 \aa 298, 39

\vfill\eject

\noindent
{\bf Figure Captions} 

\medskip
\noindent
{\bf Figure 1.} Dereddened instrumental 
CMD for $\sim 15,800$ stars in a field in the
inner disk of the LMC.
The diagonal dash indicates the reddening correction applied.
The horizontal dashed line indicates the saturation limit.

\medskip
\noindent 
{\bf Figures 2a-c.} CMDs in the Johnson-Cousins system for 
$\sim 15,800$ stars in our field.  Representative
Poisson error bars
are shown.  The 
dashes indicate the reddening correction applied (solid) and
the systematic uncertainty (dotted), as discussed in the text.  The horizontal
dashed line indicates the saturation limit.  The filled circles are 
the loci of the distribution derived from Figs. 3a-l.  
Superposed are the following isochrones: (a) A zero-age main-sequence
with $Z=0.01$ and 1,2,3 and 4 Gyr isochrones, also with $Z=0.01$;
the diagonal line illustrates our selection criterion for main-sequence
stars.
(b) A zero-age main-sequence with $Z=0.01$ and 1,2,3, and 4 Gyr
isochrones with $Z=0.004$; The box indicates the selected sample of 
Population II giants discussed in the text.
(c) A zero-age main-sequence with $Z=0.01$,
a 7 Gyr isochrone with $Z=0.001$ (dotted), and 12 and 15 Gyr
isochrones with $Z=0.0001$.  The dashed curves are the loci of the 
observed giant branches of the old LMC clusters NGC 1841 and NGC 2210.

\medskip
\noindent
{\bf Figures 3a-l.}  Histograms of $(V-I)_0$ colour constructed in 0.5
magnitude bins in $V_0$, as indicated.  
The adopted peaks (plotted as filled circles
in Figs. 2a-c) are indicated with dotted lines.

\medskip
\noindent
{\bf Figures 4a-f.} Histograms in $(V-I)_0$ in five different magnitude bins,
as indicated.  The solid histograms are for our sample, and
the dotted histograms are for a field at $\sim 4\deg$ 
in the outer disk of the LMC
studied by Gallagher \etal (1996).  The agreement in colour is excellent
in (e), while in (c) and (d), a significant offset is present.
(f) shows the comparison for the faintest magnitude bin if we had adopted
$E(B-V)=0.1$ instead of 0.2.

\medskip
\noindent
{\bf Figures 5a-c.}  The positions of the peaks in Figs. 4c and d
compared with Yale isochrones of different metallicities and ages,
as indicated.  Filled
circles are from our data and open circles are from Gallagher \etal (1996).

\medskip
\noindent
{\bf Figure 6.} Luminosity functions for main-sequence stars in our
field (solid curve), and for five other fields in the LMC from Bertelli \etal
(1992) and Vallenari \etal (1996a,b) (dotted curves). 
The luminosity functions have been normalized to
coincide at $M_V=3.0$ ($V_0=21.5$).  

\medskip
\noindent
{\bf Figures 7a, b.} Surface density of Population II giants in the
two fields studied with HST (circles), and of RRLyraes from 
Kinman \etal (1991) (triangles).
The RRLyrae data have been normalized to overlap the stellar
data.  $\sqrt N$ errorbars are shown.
(a) Two King models have core and tidal radii as indicated.
The solid curve is fit to the two HST points, and the dashed curve
is as adopted by
Kinman \etal (b) The same data plotted so that a straight line corresponds
to an exponential disk.  The dashed
line is  the fit adopted by Kinman \etal for the RRLyrae data and
has exponential scale length $\alpha \sim 0.39$ kpc$^{-1} 
= 0.34 $deg $^{-1}$.

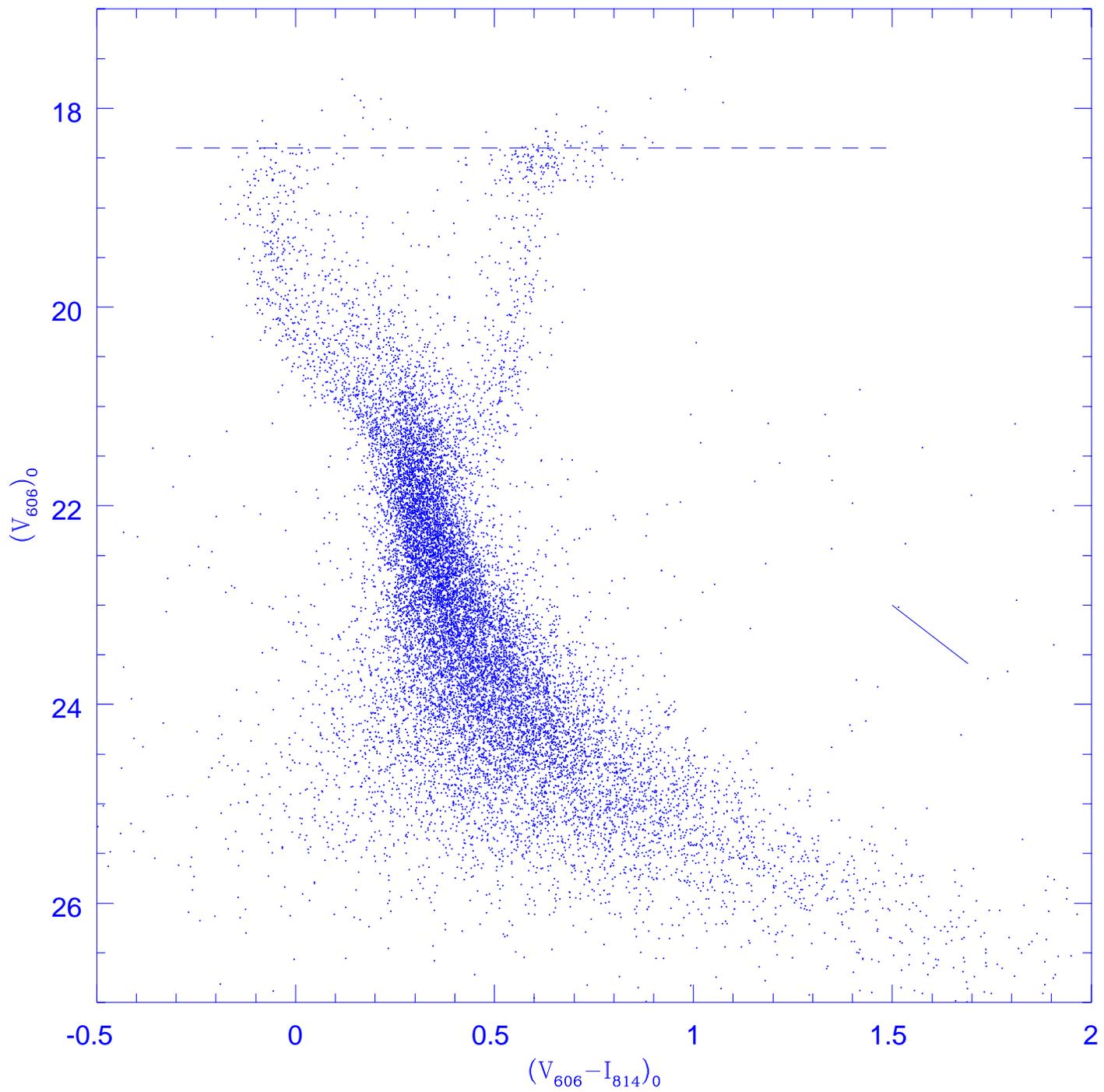
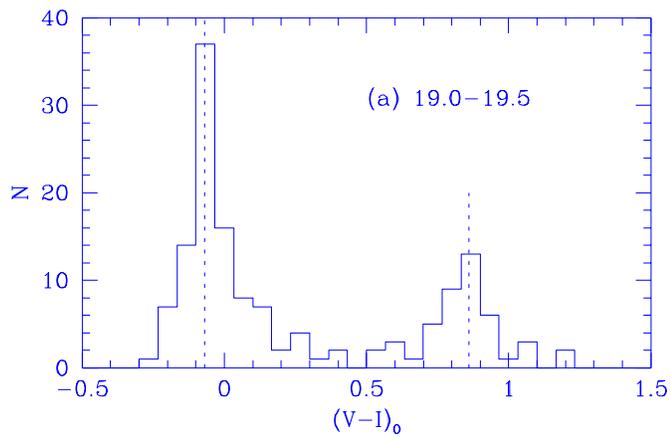
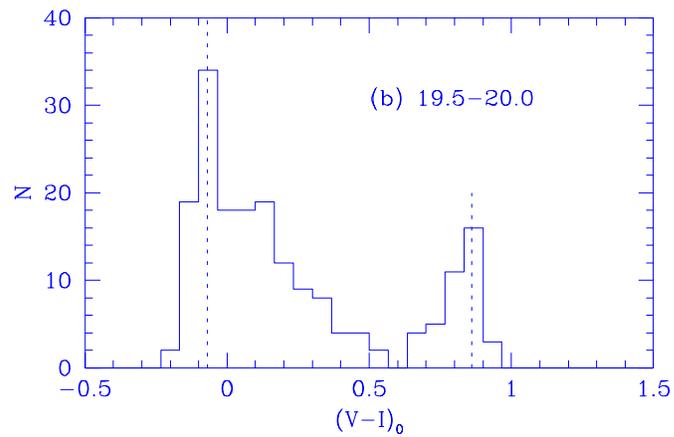
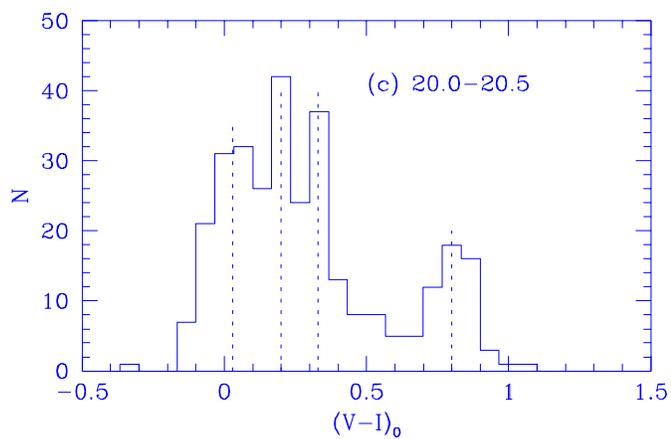
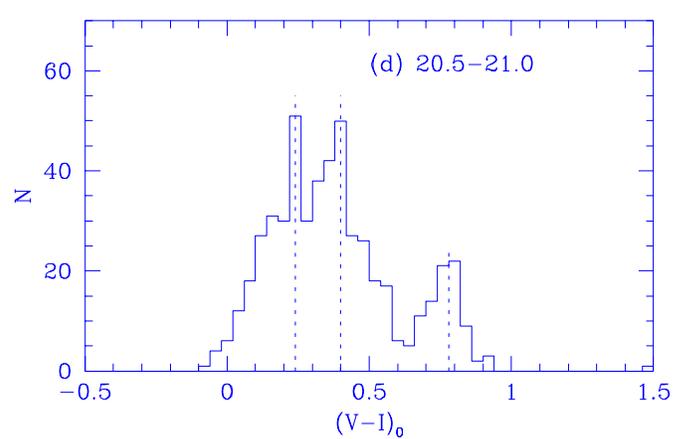
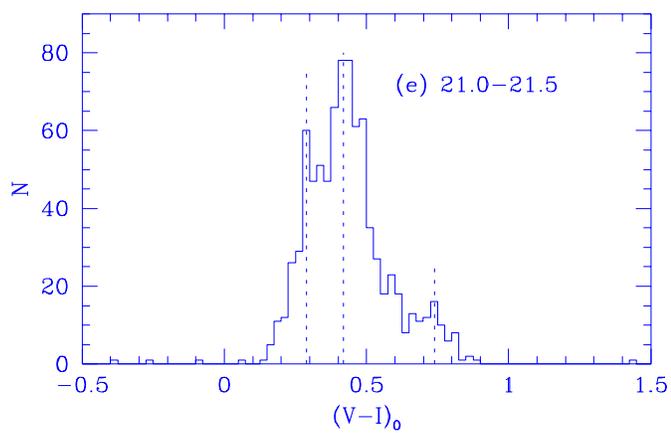
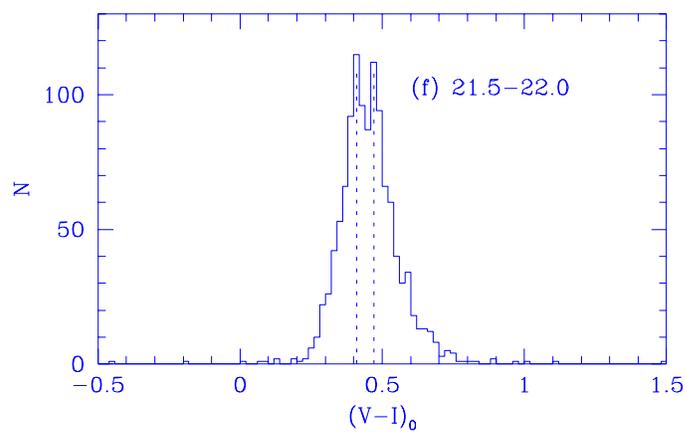
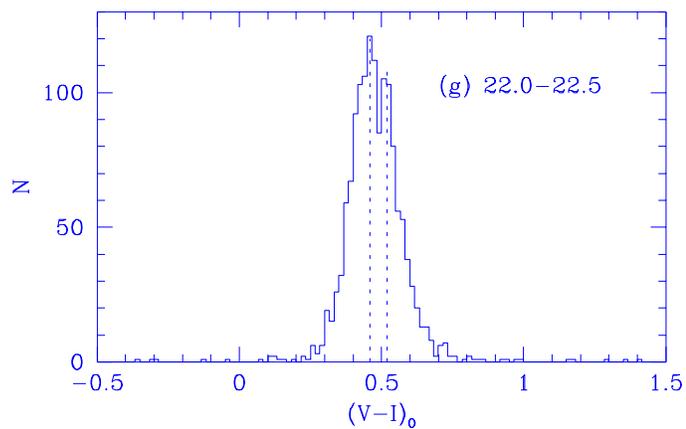
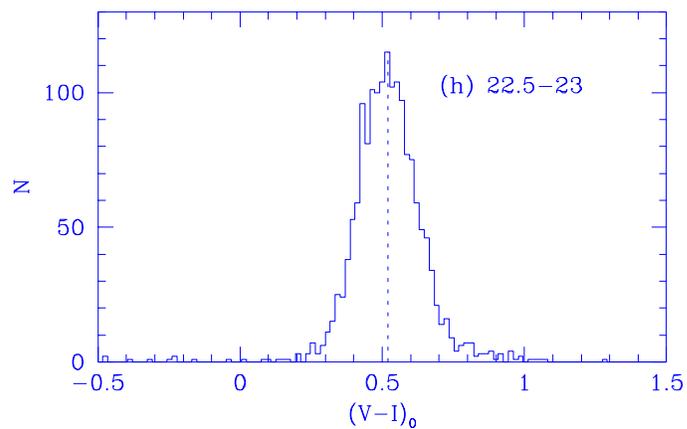
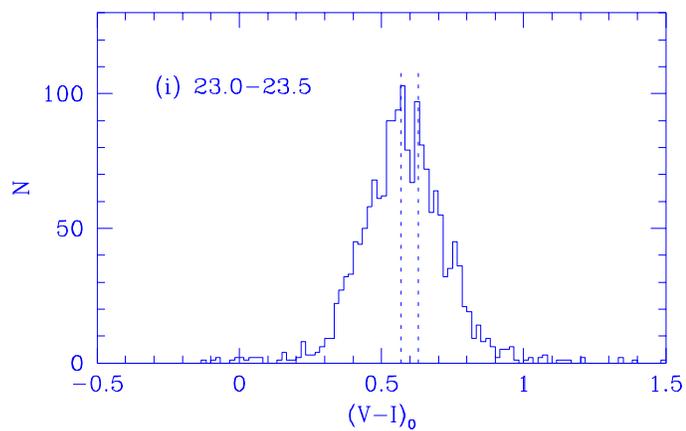
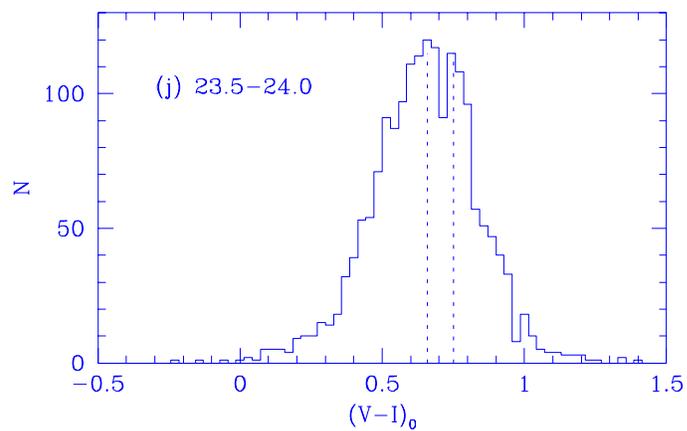
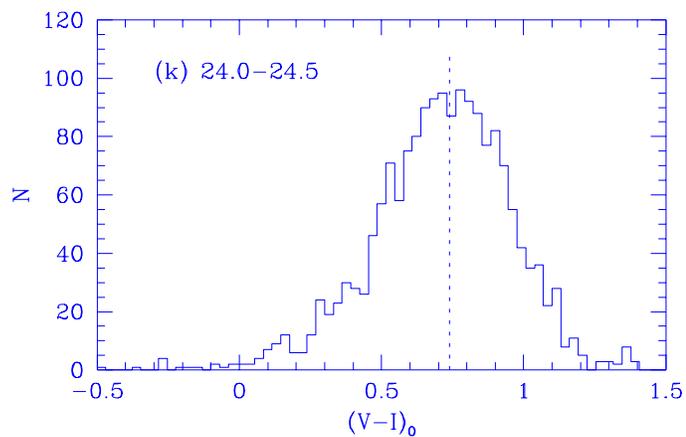
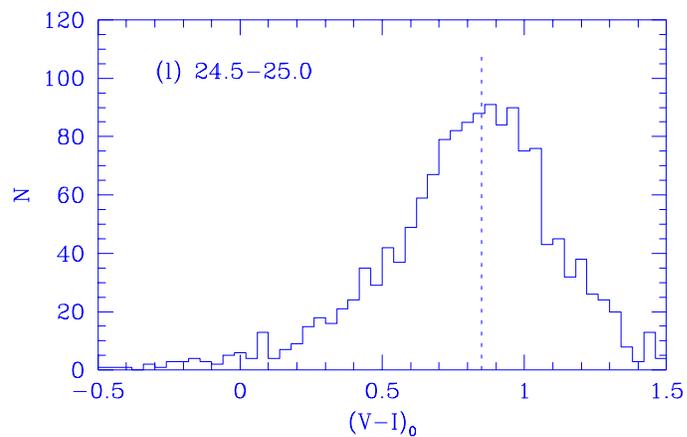
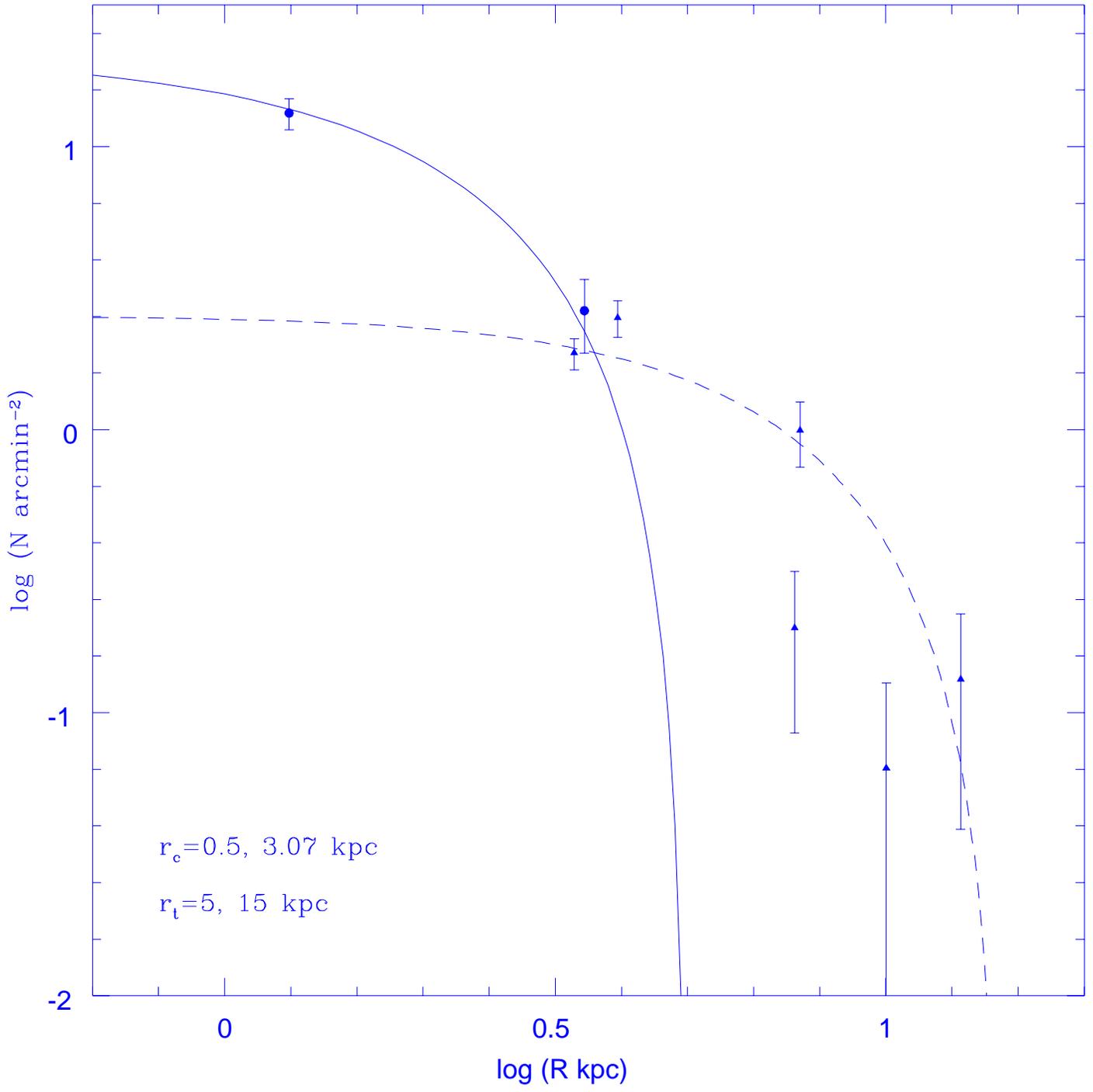
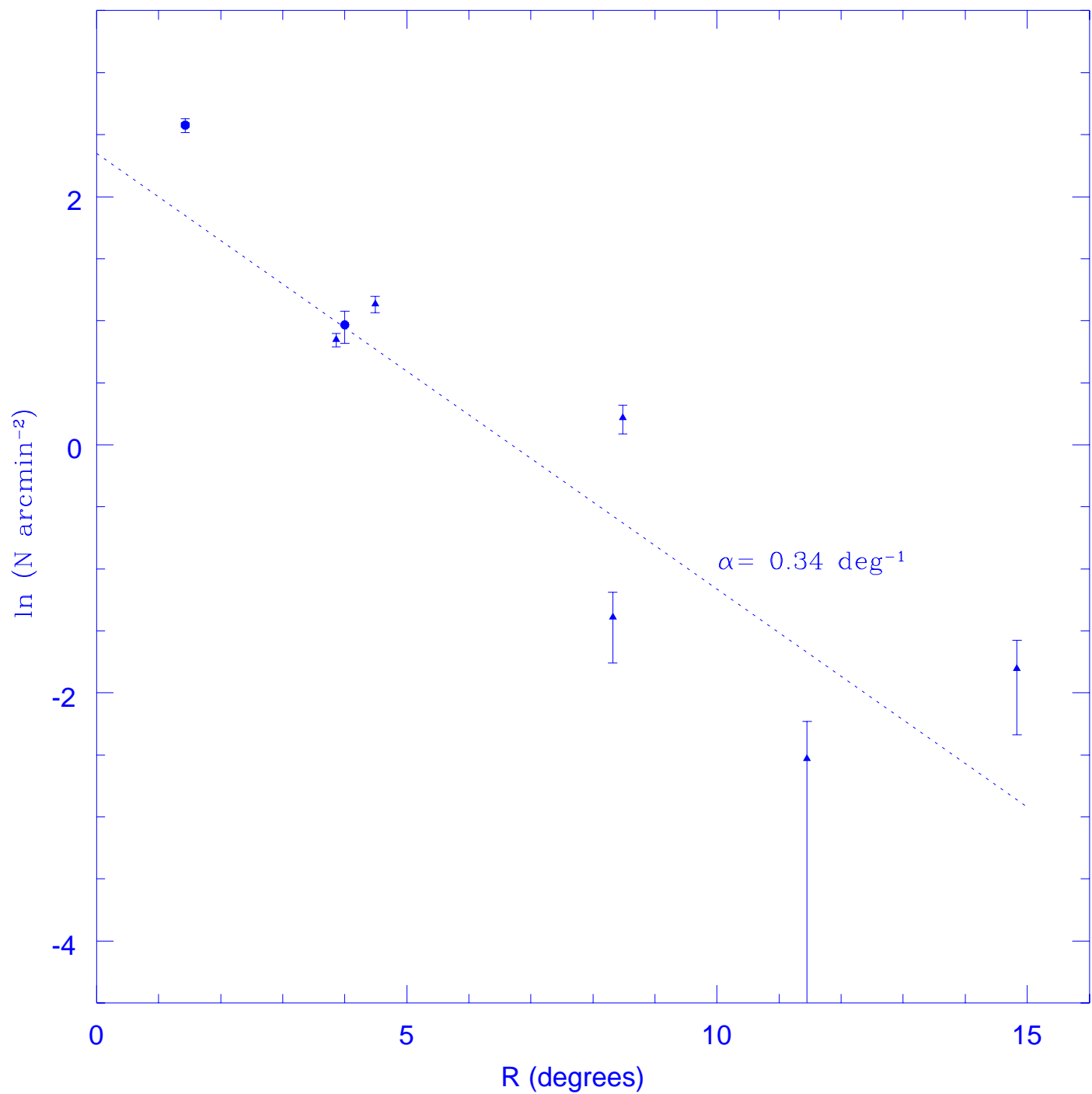
\end